\documentclass{elsart}

\usepackage{adnd}
\usepackage{longtable}

\usepackage{mathptmx}

\usepackage{amsmath}

\usepackage{amssymb,amstext}
\usepackage[pdftex,
 letterpaper=true,
 hyperindex=true,
 breaklinks=true,
 colorlinks=false,
 citecolor=blue,
 pdftitle={},
 pdfauthor={}]
{hyperref}

\usepackage{graphicx}

\begin{document}

\begin{frontmatter}

\journal{Atomic Data and Nuclear Data Tables}

\copyrightholder{Elsevier Science}

\runtitle{Krypton}
\runauthor{Heim}


\title{Discovery of the Krypton Isotopes}


\author{M. Heim},
\author{A. Fritsch},
\author{A. Schuh},
\author{A. Shore},
\and
\author{M.~Thoennessen\corauthref{cor}},\corauth[cor]{Corresponding author.}\ead{thoennessen@nscl.msu.edu}

\address{National Superconducting Cyclotron Laboratory, and \\ Department of Physics and Astronomy, Michigan State University, \\East Lansing, MI 48824, USA}

\date{March 24, 2009} 

\begin{abstract}
Thirty-two krypton isotopes have been observed so far; the discovery of these isotopes is discussed.  For each isotope a brief summary of the first refereed publication, including the production and identification method, is presented.
\end{abstract}

\end{frontmatter}





\newpage
\tableofcontents
\listofDtables

\vskip5pc

\section{Introduction}\label{s:intro}

In this fifth paper in the series of the discovery of isotopes, the discovery of the krypton isotopes is discussed. Previously, the discovery of cerium \cite{Gin09}, arsenic \cite{Sho09}, gold \cite{Sch09}, and tungsten \cite{Fri09} isotopes was discussed. The purpose of this series is to document and summarize the discovery of the isotopes. Guidelines for assigning credit for discovery are (1) clear identification, either through decay-curves and relationships to other known isotopes, particle or $\gamma$-ray spectra, or unique mass and Z-identification, and (2) publication of the discovery in a refereed journal. The authors and year of the first publication, the laboratory where the isotopes were produced as well as the production and identification methods are discussed. When appropriate, references to conference proceedings, internal reports, and theses are included. When a discovery includes a half-life measurement, the measured value is compared to the currently adapted value taken from the NUBASE evaluation \cite{Aud03}, which is based on ENSDF database \cite{ENS08}. In cases where the reported half-life differed significantly from the adapted half-life (up to approximately a factor of two), we searched the subsequent literature for indications that the measurement was erroneous. If that was not the case, we credited the authors with the discovery in spite of the inaccurate half-life.

\section{Discovery of $^{69-100}$Kr}

Thirty-two krypton isotopes from A = $69-100$ have been discovered so far; these include 6 stable, 11 proton-rich and 15 neutron-rich isotopes.  According to the HFB-14 model \cite{Gor07}, $^{115}$Kr should be the last odd particle stable neutron-rich nucleus and the even particle stable neutron-rich nuclei should continue through $^{126}$Kr. Three more neutron-deficient isotopes ($^{66-68}$Kr) are predicted to be stable. Thus, there remain 24 isotopes to be discovered. In addition, it is estimated that 2 additional nuclei beyond the proton dripline could live long enough to be observed \cite{Tho04}. Almost 60\% of all possible krypton isotopes have been produced and identified so far.

\begin{figure}
	\centering
	\includegraphics[scale=.5]{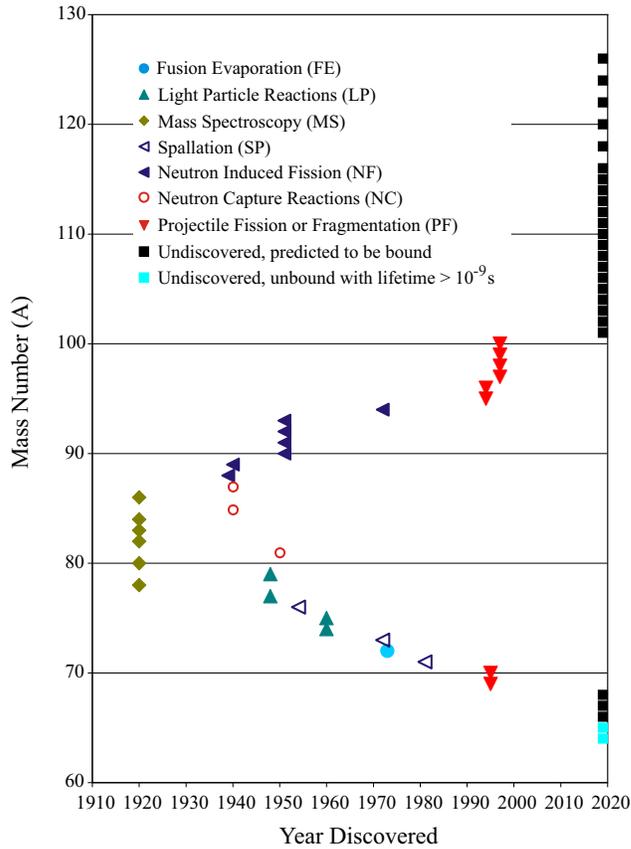}
	\caption{Krypton isotopes as a function of time they were discovered. The different production methods are indicated. The solid black squares on the right hand side of the plot are isotopes predicted to be bound by the HFB-14 model.  On the proton-rich side, the light blue squares correspond to unbound isotopes predicted to have lifetimes larger than $\sim 10^{-9}$~s.}
\label{f:year}
\end{figure}

Figure \ref{f:year} summarizes the year of first discovery for all krypton isotopes identified by the method of discovery.  The range of isotopes predicted to exist is indicated on the right side of the figure. The radioactive krypton isotopes were produced using heavy-ion fusion evaporation (FE),  light-particle reactions (LP), neutron-induced fission (NF), neutron-capture reactions (NC), spallation reactions (SP), and projectile fragmentation or fission (PF). The stable isotopes were identified using mass spectroscopy (MS). Heavy ions are all nuclei with an atomic mass larger than A = 4 \cite{Gru77}. Light particles also include neutrons produced by accelerators. In the following paragraphs, the discovery of each krypton isotope is discussed in detail.

\subsection*{$^{69,70}$Kr}\vspace{-0.85cm}
Blank {\it{et al.}} reported the discovery of the isotopes $^{69}$Kr and $^{70}$Kr in {\it{New Isotopes From $^{78}$Kr Fragmentation and the Ending Point of the Astrophysical Rapid-Proton-Capture Process}} in 1995. They produced the isotopes using SISSI and LISE at the Grand Acc\'{e}l\'{e}rateur National d'Ions Lourds (GANIL) facility \cite{Bla95}. The discovery was made using projectile fragmentation with $^{78}$Kr at 73 MeV/nucleon on a nickel target. ``We obtained clear evidence for the existence of the five new isotopes $^{60}$Ga, $^{64}$As, $^{69,70}$Kr, and $^{74}$Sr.'' Identification of the new isotope was achieved using time of flight and $\Delta$E-E measurements and confirmed by measuring $\gamma$ decays of known isomers.

\subsection*{$^{71}$Kr}\vspace{-0.85cm}
Ewan {\it{et al.}} discovered $^{71}$Kr in 1981 using the Isotope Separator On Line (ISOLDE) facility at CERN, Switzerland, as reported in {\it{Beta-Decay of the T$_{z}$ = $\frac{1}{2}$ Mirror Nucleus $^{71}$Kr.}} \cite{Ewa81}. A niobium target was bombarded with 600 MeV protons, followed by mass separation. ``Assignment of the $\beta$-activity to mass 71 was made by scanning over this mass region with the separator magnet.'' The half-life was measured to be 97(9)~ms. This value is included in the average for the presently accepted value of 100(3)~ms.

\subsection*{$^{72}$Kr}\vspace{-0.85cm}
First reported in the article {\it{A New N = Z Isotope: Krypton 72}}, Schmeing {\it{et al.}} made the discovery of $^{72}$Kr in 1973 using the upgraded Chalk River MP tandem \cite{Sch73}. Production of the isotope occurred by the fusion-evaporation reaction of $^{16}$O at 55 MeV with a $^{58}$Ni target, $^{58}$Ni($^{16}$O,2n)$^{72}$Kr. The reaction products were thermalized in the helium target cell and periodically swept to a shielded counting cell where $\beta$-delayed $\gamma$-rays were observed. ``We determine the half-life of $^{72}$Kr to be 16.7$\pm$0.6~s.'' This agrees with the accepted value of 17.16(18)~s. It should be mentioned that within a month of the submission by Schmeing, Davids {\it{et al.}} submitted their results reporting the observation of $^{72}$Kr \cite{Dav73}.

\subsection*{$^{73}$Kr}\vspace{-0.85cm}
$^{73}$Kr was discovered by Hornsh\o j {\it{et al.}} in 1972 using ISOLDE at CERN and reported in {\it{Beta-Delayed Proton Emitter $^{73}$Kr}} \cite{Hor72}. The discovery was made using the spallation of a Zr(OH)$_{4}$ target by 600 MeV protons. ``The nuclide $^{73}$Kr has been identified by on-line mass separation as a precursor of $\beta$-delayed proton emission.'' The measured half-life of 34(4)~s is close to the currently accepted value of 28.8(6)~s.

\subsection*{$^{74,75}$Kr}\vspace{-0.85cm}
$^{74}$Kr and $^{75}$Kr were discovered in 1960 by Butement and Boswell and reported in {\it{New Neutron Deficient Isotopes of Krypton}} \cite{But60}. The isotopes were produced by way of bombarding a lithium bromide target with 280 MeV protons from the Liverpool Synchrocyclotron. ``The half-life of $^{74}$Kr was obtained by milking bromine from krypton at successive intervals of either 10 or 12 min and then finding the yield of $^{74}$Br by examining the decay curve of each bromine fraction counted through 750 mg/cm$^2$ of aluminium. ... The half-life of $^{75}$Kr was obtained by using a similar technique but this time bromine was milked from krypton during successive 5 min periods.'' The measured half-lives of 12(1)~m and 5.5(4)~m agree with the accepted values of 11.5(11)~m and 4.29(17)~m for $^{74}$Kr and $^{75}$Kr, respectively. It should be mentioned that Gray {\it{et al.}} discovered $^{74}$Kr essentially simultaneously \cite{Gra60}. The two papers were submitted less than a month apart. Gray was apparently not aware of the work by Butement and Boswell.

\subsection*{$^{76}$Kr}\vspace{-0.85cm}
Caretto and Wiig reported the discovery of $^{76}$Kr in 1954 in {\it{A New Neutron-Deficient Isotope of Krypton.}} \cite{Car54}. ``While investigating the spallation reactions which occurred when yttrium was bombarded with 150-, 175-, and 240-Mev protons in the Rochester cyclotron, a new krypton isotope, Kr$^{76}$, with a (9.7$\pm$0.5)-hour half-life was observed.'' The isotope was identified by measuring their radioactive decay with a NaI(Tl) phosphor and a beta-proportional counter following chemical separation. The measured half-life is near the currently adapted value of 14.8(1)~h.

\subsection*{$^{77}$Kr}\vspace{-0.85cm}
$^{77}$Kr was discovered by Woodward {\it{et al.}} in 1948 at Ohio State University. Discovery of the isotope was reported in the paper {\it{Radioactive Kr Isotopes}} \cite{Woo48}. Enriched samples of $^{74}$Se and $^{76}$Se were bombarded by $\alpha$-particles, and $\beta$$^{+}$ and $\beta$$^{-}$ radiation was measured; ``... a strong $\beta^+$-activity of 1.1-hour half-life was observed in the enriched Se$^{74}$ sample but not in the enriched Se$^{76}$ sample.'' Thus $^{77}$Kr was formed in the reaction $^{74}$Se($\alpha$,n). The measured 1.1~h half-life is consistent with the presently accepted value of 74.4(6)~m.

\subsection*{$^{78}$Kr}\vspace{-0.85cm}
The discovery of the stable isotope $^{78}$Kr was made by Aston in 1920 with a mass spectrograph, and he reported the results in {\it{Isotopes and Atomic Weights}} \cite{Ast20a}. Aston identified six separate isotopes of krypton; ``Krypton has no fewer than six constituents: 78, 80, 82, 83, 84, and 86. The last five are strong lines most beautifully confirmed by double- and triple-charged clusters, which can be compared with great accuracy against A (4) and CO (28). The 78 line has not yet been confirmed in this way owing to its faintness, but there is no reason to doubt its elemental nature. Krypton is the first element giving unmistakable isotopes differing by one unit only." Aston published the actual spectra a few months later \cite{Ast20b}.

\subsection*{$^{79}$Kr}\vspace{-0.85cm}
$^{79}$Kr was uniquely identified for the first time by Woodward {\it{et al.}} in 1948 at Ohio State University in the same paper describing the discovery of $^{77}$Kr: {\it{Radioactive Kr Isotopes}} \cite{Woo48}. Enriched samples of $^{74}$Se and $^{76}$Se were bombarded by $\alpha$-particles and $\beta$$^{+}$ and $\beta$$^{-}$ radiation was measured. ``These curves indicate that the 1.4-day krypton period is formed from stable Se$^{76}$. ... The 1.4-day krypton activity is then assigned to Kr$^{79}$.'' This half-life agrees with the currently accepted value of 35.04(10)~h. Earlier half-lives of 18(2)~h \cite{Sne37}, 34.5(10)~h \cite{Cre40}, and 33-35~h \cite{Cla41} had been observed, but it had not been possible to make the unique assignment to $^{79}$Kr. These earlier observations assigned the activity to either $^{79}$Kr or $^{81}$Kr.

\subsection*{$^{80}$Kr}\vspace{-0.85cm}
The discovery of the stable isotope $^{80}$Kr was made by Aston in 1920 with a mass spectrograph, and he reported the results in {\it{Isotopes and Atomic Weights}} \cite{Ast20a}. Aston identified six separate isotopes of krypton; ``Krypton has no fewer than six constituents: 78, 80, 82, 83, 84, and 86. The last five are strong lines most beautifully confirmed by double- and triple-charged clusters, which can be compared with great accuracy against A (4) and CO (28). ... Krypton is the first element giving unmistakable isotopes differing by one unit only." Aston published the actual spectra a few months later \cite{Ast20b}.

\subsection*{$^{81}$Kr}\vspace{-0.85cm}
Reynolds reported the discovery of $^{81}$Kr in {\it{A New Long-Lived Krypton Activity}} in 1950 at Argonne National Laboratory \cite{Rey50}. The discovery was made by distilling krypton from a sample of sodium bromide, which had gone through prolonged exposure to intense neutron radiation. The isotope was first noticed due to an ion bump at mass 81 and, upon further inspection, was identified to be $^{81}$Kr. ``In addition, a small ion peak was observed at mass 81 which, after the usual tests common to mass spectrometric technique, proved to be due to Kr$^{81}$ and not due to an impurity element or to a rare krypton compound ion such as KrH$^{+}$." Reynolds concluded $^{81}$Kr decays by way of K-capture with a half-life of 2.1(5)$\times$10$^5$~y. This half-life agrees with the adapted value of 2.29(11)$\times$10$^5$~y. The 13.1~s isomers had been observed by Creutz {\it{et al.}} in 1940; however, they could assign the observed 13(2)~s activity only to either $^{79}$Kr or $^{81}$Kr \cite{Cre40}.

\subsection*{$^{82-84}$Kr}\vspace{-0.85cm}
The discovery of the stable isotopes $^{82}$Kr, $^{83}$Kr, and $^{84}$Kr, was made by Aston in 1920 with a mass spectrograph, and he reported the results in {\it{Isotopes and Atomic Weights}} \cite{Ast20a}. Aston identified six separate isotopes of krypton; ``Krypton has no fewer than six constituents: 78, 80, 82, 83, 84, and 86. The last five are strong lines most beautifully confirmed by double- and triple-charged clusters, which can be compared with great accuracy against A (4) and CO (28). ... Krypton is the first element giving unmistakable isotopes differing by one unit only." Aston published the actual spectra a few months later \cite{Ast20b}.

\subsection*{$^{85}$Kr}\vspace{-0.85cm}
In 1943, Born and Seelmann-Eggebert were the first to identify $^{85}$Kr in Berlin in their paper {\it{\"Uber die Identifizierung einiger Uranspaltprodukte mit entsprechenden durch (n$\alpha$)- und (np)-Prozesse erhaltenen Isotopen}} \cite{Bor43}. Rubidium and strontium salts were irradiated with neutrons from the high-voltage facility of the Kaiser Wilhelm Institut f\"ur Physik and decay curves following chemical separation were measured. ``Von diesen sind 83 und 84 stabil, so da\ss\ dem 4.6-Std.-Krypton offenbar die Masse 85 zuzuordnen ist. Der Bildung dieses Isotops aus Rubidium w\"urde dann dem Proze\ss\ $^{85}$Rb(np)$^{85}$Kr entsprechen.'' (Out of these, 83 and 84 are stable, so that the 4.6~h krypton obviously has to be assigned to mass 85. The production of this isotope from rubidium would thus correspond to the reaction $^{85}$Rb(n,p)$^{85}$Kr.) The half-life of 4.6~h agrees with the accepted value of 4.480(8)~h and corresponds to an isomer. A month earlier the authors detected the 4.6~h activity in the neutron-induced fission of uranium without assigning it to a specific mass \cite{See43}. Even earlier, in 1937, Snell \cite{Sne37} had also measured a 4.5~h activity, but could not assign it to a specific isotope: ``The strong 74-minute and 4.5-hour activities emit negatives, which shows that they probably belong to krypton 85 and 87.'' It also should be mentioned that Clancy had argued to assign the ($^{85}$Kr) 4~h activity to $^{87}$Kr \cite{Cla41,Cla40}.

\subsection*{$^{86}$Kr}\vspace{-0.85cm}
The discovery of the stable isotope $^{86}$Kr was made by Aston in 1920 with a mass spectrograph and he reported the results in {\it{Isotopes and Atomic Weights}} \cite{Ast20a}. Aston identified six separate isotopes of krypton; ``Krypton has no fewer than six constituents: 78, 80, 82, 83, 84, and 86. The last five are strong lines most beautifully confirmed by double- and triple-charged clusters, which can be compared with great accuracy against A (4) and CO (28). ... Krypton is the first element giving unmistakable isotopes differing by one unit only." Aston published the actual spectra a few months later \cite{Ast20b}.

\subsection*{$^{87}$Kr}\vspace{-0.85cm}
In 1943 Born and Seelmann-Eggebert were the first to identify $^{87}$Kr in Berlin in their paper {\it{\"Uber dir Identifizierung einiger Uranspaltprodukte mit entsprechenden durch (n$\alpha$)- und (np)-Prozesse erhaltenen Isotopen}} \cite{Bor43}. Rubidium and strontium salts were irradiated with neutrons from the high-voltage facility of the Kaiser Wilhelm Institut f\"ur Physik and decay curves following chemical separation were measured. ``Widerspruchslos l\"a\ss t sich unter diesen Voraussetzungen das 75-Min.-Krypton der Masse 87 zuordnen.'' (Without objections, the 75~m krypton can be assigned under these circumstances to mass 87.) The half-life of 75~m agrees with the accepted value of 76.3(5)~m. A month earlier the authors detected the 75~m activity in the neutron-induced fission of uranium without assigning it to a specific mass \cite{See43}. Even earlier, in 1937, Snell \cite{Sne37} had also measured a 74(2)~m activity, but could not assign it to a specific isotope: ``The strong 74-minute and 4.5-hour activities emit negatives, which shows that they probably belong to krypton 85 and 87.'' It also should be mentioned that Clancy had argued to assign the ($^{85}$Kr) 4~h activity to $^{87}$Kr \cite{Cla41,Cla40}.

\subsection*{$^{88}$Kr}\vspace{-0.85cm}
Langsdorf discovered $^{88}$Kr in 1939 at the Radiation Laboratory at Berkeley and reported his findings in {\it{Fission Products of Thorium}} \cite{Lan39}. ``Several long-lived noble gases from thorium irradiated with fast neutrons (9 MeV, from deuterons on beryllium) have been observed. One of these gases is a krypton of 3-hour half-life which decays into an 18-minute rubidium...'' The 3~h half-life is consistent with the presently accepted value of 2.84(3)~h. The existence of $^{88}$Kr had been noted three months earlier by Heyn {\it{et al.}} \cite{Hey39}. However, Heyn only measured the subsequent decay of $^{88}$Rb and inferred that krypton ($^{88}$Kr) had escaped from the irradiated uranium solution.

\subsection*{$^{89}$Kr}\vspace{-0.85cm}
Hahn and Strassmann reported the first identification of $^{89}$Kr in Berlin in the 1943 paper  {\it{\"Uber die bei der Uranspaltung auftretenden aktiven Strontium- und Yttrium-Isotope}} \cite{Hah43a}. The isotope was observed following neutron irradiation of uranium and a half-life of 2.5~m was measured. ``Sicher ist nur, da\ss\ das 2.5-Minuten-Krypton \"uber ein 15.4-Minuten-Rubidium in das 55-Tage-Strontium \"ubergeht.'' (The only certain assignment is the decay of the 2.5~m krypton, via the 15.4~m rubidium to the 55~d strontium.) Further details were discussed in a subsequent publication \cite{Hah43b}. The half-life is close to the presently accepted value of 3.15(4)~m. A krypton activity of 2.5$-$3~m had already been observed in 1940 by Seelman-Eggebert \cite{See40}; however, it could only be linked to a 15.5~m rubidium activity which had been reported by Glasoe and Steigman who at that time had not assigned a mass to the activity \cite{Gla40a}. Only a few weeks later Glasoe and Steigman \cite{Gla40b} assigned the decay to $^{89}$Rb and indirectly implied the existence of $^{89}$Kr unaware of the Seelman-Eggebert result: ``From the manner in which this Rb activity is obtained it is estimated that the Kr parent must have a period of the order of one to five minutes.''

\subsection*{$^{90}$Kr}\vspace{-0.85cm}
Kofoed-Hansen and Nielsen reported the discovery of  $^{90}$Kr in 1951 at Copenhagen, Denmark, in {\it{Short-Lived Krypton Isotopes and Their Daughter Substances}} \cite{Kof51}. ``Krypton formed in fission of uranium was pumped through a 10-m long tube directly form the cyclotron into the ion source of the isotope separator.'' Gamma- and $\beta^-$-radiation was measured and a half-life of 33~s was determined. This value agrees with the currently accepted value of 32.32(9)~s. Dillard {\it{et al.}} reported in two papers of the Manhattan Project Technical Series only a short half-life \cite{Dil51b} and an estimated half-life of 25~s \cite{Dil51a}. The previously reported half-life of 33~s mentioned by Kofoed-Hansen and Nielsen was only published in an internal report \cite{Kat46}.

\subsection*{$^{91-93}$Kr}\vspace{-0.85cm}
Dillard {\it{et al.}} from Argonne National Laboratory reported the discovery of $^{91}$Kr, $^{92}$Kr, and $^{93}$Kr in 1951 as part of the Manhattan Project Technical Series: {\it{Determination of Gas Half-Life By The Charged-Wire Technique (II)}} \cite{Dil51a}. ``The active isotopes of krypton and xenon produced in neutron-irradiated uranium have been investigated by the charged-wire collection technique.''
The measured half-life for $^{91}$Kr of  9.8(5)~s is close to the accepted value of 8.57(4)~s. In two other papers of this technical series the half-life was estimated to be 6~s \cite{Dil51b} and 5.7~s \cite{Ove51}. It should be mentioned that in February 1951 Kofoed-Hansen and Nielsen \cite{Kof51} reported a half-life of 10~s for $^{91}$Kr. The authors were aware of the results of the Manhattan Project.
The observed half-life for $^{92}$Kr of 3.0(5)~s, which was tentatively assigned to $^{92}$Kr is close to the currently accepted value of 1.840(8)~s.
In the main text of reference \cite{Dil51a} the reported half-life of 2.0(5)~s for $^{93}$Kr was assigned to $^{95}$Kr. However, the correct mass of 83 is assigned in a footnote referring to another paper of the Manhattan Project Technical Series \cite{Sel51}. In addition, in a different paper of the series the authors had assigned an estimated half-life of 1-2~s correctly to $^{93}$Kr \cite{Dil51b}. The quoted half-life is close to the accepted value of 1.286(10)~s.

\subsection*{$^{94}$Kr}\vspace{-0.85cm}
Amiel {\it{et al.}} discovered $^{94}$Kr in 1972 at the Soreq Nuclear Research Centre, Israel, as reported in {\it{Identification of $^{94}$Kr and $^{143}$Xe, and Measurement of $\gamma$-Ray Spectra and Half-Lives of Nuclides in the Mass-Chains 93, 94, and 143}} \cite{Ami72}. ``The target was irradiated in a thermal-neutron flux of approximately 2$\times$10$^8$ n cm$^{-2}$ sec$^{-1}$ from an external beam tube of the Israel Research Reactor-1.'' The target contained 9~g of U$_3$O$_8$ of 93.3\% enriched $^{235}$U. $^{94}$Kr was identified by $\beta$-decay curves measured at the Soreq On-Line Isotope Separator. The measured half-life of 0.20(1)~s agrees with the currently adapted value of 0.210(4)s. It should be mentioned that a half-life of 1.4(5)~s was reported by Dillard {\it{et al.}} as the krypton parent of a 20~m yttrium isotope. Although this isotope is most likely $^{94}$Y (T$_{1/2}$ = 18.7(1)~m), Dillard {\it{et al.}} is not credited with the discovery because they were not able to specify the mass directly. In addition, the measured half-life is significantly larger than the accepted value. We also do not credit the determination of fractional fission yields of $^{94}$Kr from nuclear charge distribution measurements in low-energy fission \cite{Wah62} because it was not directly identified and the half-life was not measured.

\subsection*{$^{95}$Kr}\vspace{-0.85cm}
The credit for the discovery of $^{95}$Kr is attributed to Bernas {\it{et al.}} in the 1994 paper {\it{Projectile Fission at Relativistic Velocities: A Novel and Powerful Source of Neutron-Rich Isotopes Well Suited for In-Flight Isotopic Separation}} \cite{Ber94}. A 750 MeV/nucleon $^{238}$U beam accelerated by the GSI UNILAC-SIS accelerator system was used to produce $^{95}$Kr in projectile fission on a lead target. The authors do not mention the discovery of $^{95}$Kr because its existence had previously been reported. The fractional fission yields of $^{95}$Kr were determined from nuclear charge distribution measurements in low-energy fission \cite{Wah62}; however, these measurements are not credited with the discovery because $^{95}$Kr was not directly identified and the half-life was not measured. In 1976, Ahrens {\it{et al.}} extracted the half-life of $^{95}$Kr (T$_{1/2}$ = 0.78(3)~s) from its long-lived decay products using a gas-flow method \cite{Ahr76}. However, in 2003 Bergmann {\it{et al.}} \cite{Ber03} measured a significantly shorter half-life of 114(3)~ms which raises doubt about the Ahrens measurement: ``In particular, the half-lives from the
earlier indirect radiochemical measurements ... (quoted by nuclear data evaluators for $^{95}$Kr ...) deviate considerably from our results, indicating that these identifications probably were not correct.'' Thus, we credit the discovery to Bernas {\it{et al.}} who produced $^{95}$Kr after Ahrens {\it{et al.}} but prior to Bergman {\it{et al.}} However, this issue warrants further evaluation.

\subsection*{$^{96}$Kr}\vspace{-0.85cm}
Bernas {\it{et al.}} discovered $^{96}$Kr in 1994 at GSI, Germany, as reported in {\it{Projectile Fission at Relativistic Velocities: A Novel and Powerful Source of Neutron-Rich Isotopes Well Suited for In-Flight Isotopic Separation}} \cite{Ber94}. The isotope was produced using projectile fission of $^{238}$U at 750 MeV/nucleon on a lead target. ``Forward emitted fragments from $^{80}$Zn up to $^{155}$Ce were analyzed with the Fragment Separator (FRS) and unambiguously identified by their energy-loss and time-of-flight.'' The experiment yielded 155 individual counts of $^{96}$Kr.

\subsection*{$^{97-100}$Kr}\vspace{-0.85cm}
$^{97}$Kr, $^{98}$Kr, $^{99}$Kr, and $^{100}$Kr were discovered by Bernas {\it{et al.}} in 1997 at GSI in Germany and reported in {\it{Discovery and Cross-Section Measurement of 58 New Fission Products in Projectile-Fission of 750$\cdot$AMeV $^{238}$U}} \cite{Ber97}. The experiment was performed using projectile fission of $^{238}$U at 750~MeV/nucleon on a beryllium target. ``Fission fragments were separated using the fragment separator FRS tuned in an achromatic mode and identified by event-by-event measurements of $\Delta$E-B$\rho$-ToF and trajectory.'' During the experiment, individual counts for $^{97}$Kr (2110), $^{98}$Kr (525), $^{99}$Kr (32), and $^{100}$Kr (3) were recorded. It should be mentioned that an estimated half-life of 1-2~s for $^{97}$Kr had been reported during the Manhattan Project \cite{Dil51a}. However, this observation was later questioned \cite{Wah62}, which is supported by the most recent half-life measurement of 68(7)~ms \cite{Ber03}.

\section{Summary}
The activity of five krypton isotopes ($^{79}$Kr, $^{81}$Kr, $^{85}$Kr, and $^{94}$Kr) was measured before they could be assigned to the specific isotopes. The half-life of $^{95}$Kr was accepted to be 780~ms until 27 years later, when it was measured to be significantly shorter (114~ms). Based on this large difference a reassignment of the discovery seemed justified. However, the origin of the measured long half-life should be investigated and the new half-life of $^{95}$Kr should be independently confirmed. It is also interesting to note that Krypton was the first element for which adjacent stable isotopes were identified.

\ack

This work was supported by the National Science Foundation under grants No. PHY06-06007 (NSCL) and PHY07-54541 (REU). MH was supported by NSF grant PHY05-55445.


\newpage

\section*{EXPLANATION OF TABLE}\label{sec.eot}
\addcontentsline{toc}{section}{EXPLANATION OF TABLE}

\renewcommand{\arraystretch}{1.0}

\begin{tabular*}{0.95\textwidth}{@{}@{\extracolsep{\fill}}lp{5.5in}@{}}
\textbf{TABLE I.}
	& \textbf{Discovery of Krypton Isotopes }\\
\\

Isotope & Krypton isotope \\
First Author & First author of refereed publication \\
Journal & Journal of publication \\
Ref. & Reference \\
Method & Production method used in the discovery: \\
 & FE: fusion evaporation \\
 & LP: light-particle reactions (including neutrons) \\
 & MS: mass spectroscopy \\
 & NC: neutron-capture reactions \\
 & NF: neutron-induced fission \\
 & SP: spallation \\
 & PF: projectile fragmentation or projectile fission \\
Laboratory & Laboratory where the experiment was performed\\
Country & Country of laboratory\\
Year & Year of discovery \\
\end{tabular*}
\label{tableI}

\newpage
\datatables

\setlength{\LTleft}{0pt}
\setlength{\LTright}{0pt}


\setlength{\tabcolsep}{0.5\tabcolsep}

\renewcommand{\arraystretch}{1.0}


\begin{longtable}[c]{%
@{}@{\extracolsep{\fill}}r@{\hspace{5\tabcolsep}} llllllll@{}}
\caption[Discovery of Krypton Isotopes]%
{Discovery of Krypton isotopes}\\[0pt]
\caption*{\small{See page \pageref{tableI} for Explanation of Tables}}\\
\hline
\\[100pt]
\multicolumn{8}{c}{\textit{This space intentionally left blank}}\\
\endfirsthead
Isotope & First Author & Journal & Ref. & Method & Laboratory & Country & Year \\
$^{69}$Kr & B. Blank & Phys. Rev. Lett. & Bla95 & PF & GANIL & France &1995 \\
$^{70}$Kr & B. Blank & Phys. Rev. Lett. & Bla95 & PF & GANIL & France &1995 \\
$^{71}$Kr & G.T. Ewan & Nucl. Phys. A & Ewa81 & SP & CERN & Switzerland &1981 \\
$^{72}$Kr & H. Schmeing & Phys. Lett. B & Sch73 & FE & Chalk River & Canada &1973 \\
$^{73}$Kr & P. Hornshoj & Nucl. Phys. A & Hor72 & SP & CERN & Switzerland &1972 \\
$^{74}$Kr & F.D.S. Butement & J. Inorg. Nucl. Chem. & But60 & LP & Liverpool & England &1960 \\
$^{75}$Kr & F.D.S. Butement & J. Inorg. Nucl. Chem. & But60 & LP & Liverpool & England &1960 \\
$^{76}$Kr & A.A. Caretto& Phys. Rev. & Car54 & SP & Rochester & USA &1954 \\
$^{77}$Kr & L.L. Woodward & Phys. Rev. & Woo48 & LP & Ohio State & USA &1948 \\
$^{78}$Kr & F.W. Aston & Nature & Ast20 & MS & Cavendish & UK &1920 \\
$^{79}$Kr & L.L. Woodward & Phys. Rev. & Woo48 & LP & Ohio State & USA &1948 \\
$^{80}$Kr & F.W. Aston & Nature & Ast20 & MS & Cavendish & UK &1920 \\
$^{81}$Kr & J.H. Reynolds & Phys. Rev. Lett. & Rey50 & NC & Argonne & USA &1950 \\
$^{82}$Kr & F.W. Aston & Nature & Ast20 & MS & Cavendish & UK &1920 \\
$^{83}$Kr & F.W. Aston & Nature & Ast20 & MS & Cavendish & UK &1920 \\
$^{84}$Kr & F.W. Aston & Nature & Ast20 & MS & Cavendish & UK &1920 \\
$^{85}$Kr & H.J. Born & Naturwiss. & Bor43 & NC & Berlin & Germany &1940 \\
$^{86}$Kr & F.W. Aston & Nature & Ast20 & MS & Cavendish & UK &1920 \\
$^{87}$Kr & H.J. Born & Naturwiss. & Bor43 & NC & Berlin & Germany &1940 \\
$^{88}$Kr & A. Langsdorf & Phys. Rev. Lett. & Lan39 & NF & Berkeley & USA &1939 \\
$^{89}$Kr & O. Hahn & Naturwiss. & Hah43 & NF & Berlin & Germany &1940 \\
$^{90}$Kr & O. Kofoed & Phys. Rev. Lett. & Kof51 & NF & Copenhagen & Denmark &1951 \\
$^{91}$Kr & C.R. Dillard & Nat. Nucl. Ener. Ser. & Dil51 & NF & Argonne & USA &1951 \\
$^{92}$Kr & C.R. Dillard & Nat. Nucl. Ener. Ser. & Dil51 & NF & Argonne & USA &1951 \\
$^{93}$Kr & C.R. Dillard & Nat. Nucl. Ener. Ser. & Dil51 & NF & Argonne & USA &1951 \\
$^{94}$Kr & S. Amiel & Phys. Rev. C & Ami72 & NF & Soreq & Israel &1972 \\
$^{95}$Kr & M. Bernas & Phys. Lett. B & Ber94 & PF & Darmstadt & Germany &1994 \\
$^{96}$Kr & M. Bernas & Phys. Lett. B & Ber94 & PF & Darmstadt & Germany &1994 \\
$^{97}$Kr & M. Bernas & Phys. Lett. B & Ber97 & PF & Darmstadt & Germany &1997 \\
$^{98}$Kr & M. Bernas & Phys. Lett. B & Ber97 & PF & Darmstadt & Germany &1997 \\
$^{99}$Kr & M. Bernas & Phys. Lett. B & Ber97 & PF & Darmstadt & Germany &1997 \\
$^{100}$Kr & M. Bernas & Phys. Lett. B & Ber97 & PF & Darmstadt & Germany &1997 \\

\end{longtable}

\newpage


\normalsize

\begin{theDTbibliography}{1956He83}
\bibitem[Ami72]{Ami72t} S. Amiel, H. Feldstein, M. Oron, and E. Yellin, Phys. Rev. C {\bf 5}, 270 (1972)
\bibitem[Ast20]{Ast20t} F.W. Aston, Nature {\bf 105}, 8 (1920)
\bibitem[Ber94]{Ber94t} M. Bernas, S. Czajkowski, P. Armbruster, H. Geissel, Ph. Dessagne, C. Donzaud, H-R. Faust, E. Hanelt, A. Heinz, M. Hesse, C. Kozhuharov, Ch. Miehe, G. M\"unzenberg, M. Pf\"utzner, C. R\"ohl, K.-H. Schmidt, W. Schwab, C. St\'ephan, K. S\"ummerer, L. Tassan-Got, and B. Voss, Phys. Lett. B {\bf 331}, 19 (1994)
\bibitem[Ber97]{Ber97t} M. Bernas, C. Engelmann, P. Armbruster, S. Czajkowski, F. Ameil, C. B\"ockstiegel, Ph. Dessagne, C. Donzaud, H. Geissel, A. Heinz, Z. Janas, C. Kozhuharov, Ch. Mieh\'e, G. M\"unzenberg, M. Pf\"utzner, W. Schwab, C. St\'ephan, K. S\"ummerer, L. Tassan-Got, and B. Voss, Phys. Lett. B {\bf 415}, 111 (1997)
\bibitem[Bla95]{Bla95t} B. Blank, S. Andriamonje, S. Czajkowski, F. Davi, R. Del Moral, J.P. Dufour, A. Fleury, A. Musquere, M.S. Pravikoff, R. Grzywacz, Z. Janas, M. Pfutzner, A. Grewe, A. Heinz, A. Junghans, M. Lewitowicz, J.-E. Sauvestre, and C. Donzaud, Phys. Rev. Lett. {\bf 74}, 4611 (1995)
\bibitem[Bor43]{Bor43t} H.J. Born and W. Seelmann-Eggebert, Naturwiss. {\bf 31}, 86 (1943)
\bibitem[But60]{But60t} F.D.S. Butement and G.G.J. Boswell, J. Inorg. Nucl. Chem. {\bf 16}, 10 (1960)
\bibitem[Car54]{Car54t} A.A. Caretto, JR. and E.O. Wiig, Phys. Rev. {\bf 93}, 175 (1954)
\bibitem[Dil51]{Dil51t} C.R. Dillard, R.M. Adams, H. Finston, and A. Turkevich, {\it Radiochemical Studies: The Fission Products}, Paper 68, p. 624, National Nuclear Energy Series IV, 9, (McGraw-Hill, New York 1951)
\bibitem[Ewa81]{Ewa81t} G. T. Ewan, E. Hagberg, P.G. Hansen, B. Jonson, S. Mattsson, H.L. Ravn, and P. Tidemand-Petersson, Nucl. Phys. A {\bf 352}, 13 (1981)
\bibitem[Hah43]{Hah43t} O. Hahn and F. Strassmann, Naturwiss. {\bf 31}, 249 (1943)
\bibitem[Hor72]{Hor72t} P. Hornsh{\o}j, K. Wilsky, P.G. Hansen, and B. Jonson, Nucl. Phys. A {\bf 187}, 637 (1972)
\bibitem[Kof51]{Kof51t} O. Kofoed-Hansen and K.O. Nielsen, Phys. Rev. {\bf 82}, 96 (1951)
\bibitem[Lan39]{Lan39t} A. Langsdorf, JR., Phys. Rev. Lett. {\bf 52}, 205 (1939)
\bibitem[Rey50]{Rey50t} J.H. Reynolds, Phys. Rev. Lett. {\bf }, 886 (1950)
\bibitem[Sch73]{Sch73t} H. Schmeing, J.C. Harding, R.L. Graham, J.S. Geiger, and K.P. Jackson, Phys. Lett. {\bf 44B}, 449 (1973)
\bibitem[Woo48]{Woo48t} L.L. Woodward, D.A. McCown, and M.L. Pool, Phys. Rev. {\bf 74}, 761 (1948)

\end{theDTbibliography}

\end{document}